\documentclass[prd,final,superscriptaddress,twocolumn,authoryear,nofootinbib]{revtex4-2}
\usepackage{amssymb}
\usepackage{lipsum}
\usepackage{orcidlink}
\usepackage{longtable}
\usepackage{tabularray}

\begin{document}


\title{A Search for Binary Black Hole Mergers in LIGO O1--O3 Data with Convolutional Neural Networks}

\author{Ethan Silver\orcidlink{0000-0002-1804-3960}}
\affiliation{Department of Physics, Harvard University, 17 Oxford Street Cambridge, MA 02138, USA}
\affiliation{The NSF AI Institute for Artificial Intelligence and Fundamental Interactions, USA}
\author{Plamen Krastev\orcidlink{0000-0002-0830-1479}}
\affiliation{The NSF AI Institute for Artificial Intelligence and Fundamental Interactions, USA}
\affiliation{Faculty of Arts and Sciences Research Computing, Harvard University, 38 Oxford Street, Cambridge, MA 02138, USA}
\author{Edo Berger\orcidlink{0000-0002-9392-9681}}
\affiliation{The NSF AI Institute for Artificial Intelligence and Fundamental Interactions, USA}
\affiliation{Center for Astrophysics | Harvard \& Smithsonian, 60 Garden Street, Cambridge, MA 02138-1516, USA}

\begin{abstract}
Since the first detection of gravitational waves in 2015 by LIGO from the binary black hole merger GW150914, gravitational-wave astronomy has developed significantly, with over 200 compact binary merger events cataloged. The use of neural networks has the potential to significantly speed up the detection, classification, and especially parameter estimation for gravitational wave events, compared to current techniques, quite important for electromagnetic follow-up of events. In this work, we present a machine learning pipeline using neural networks to detect gravitational wave events. We generate training data using real LIGO data to train and refine neural networks that can detect binary black hole (BBH) mergers, and apply these models to search through LIGO’s first three observing runs. We detect 57 out of the 75 total cataloged BBH events with two detectors of data in O1, O2, and O3, with 57 false positives that can mostly be ruled out with parameter inference and human inspection.
Finally, we extensively test this pipeline on time-shifted data to characterize its False Alarm Rate (FAR). These results are an important step in developing machine learning-based GW searches, enabling low-latency detection and multi-messenger astronomy.
\end{abstract}

\maketitle

\section{Introduction}
\label{introduction}

Gravitational-wave (GW) astronomy has opened up a new way of observing and exploring the universe. Following the first detection of gravitational waves from a binary black hole (BBH) merger \citep{Abbott_2016} by the two Laser Interferometer Gravitational-Wave Observatory (LIGO) detectors \citep{Aasi_2015} in 2015, there are now over 200 compact binary coalescence (CBC) detections reported by the LIGO-Virgo-KAGRA (LVK) collaboration during the first four observing runs, which include a total of 51 months from 2015--2025. Due to the increasing sensitivity of the interferometers, the number of CBC detections has increased in each observing run, with 3 detections during O1, 8 detections during O2, 79 detections during O3, and over 200 detections during the first 23 months of the recently-completed O4 \citep{Abbott_2023b, LIGO_2025a}.  While most of the detected events are BBH mergers, they also include a small number of binary neutron star (BNS) and neutron star-black hole (NSBH) mergers \citep{Abbott_2017a,Abbott_2020,Abbott_2021b,LIGO_2025b,LIGO_2025c}. With future ground- and space-based detectors, the number of detections is expected to increase dramatically, paving the way for GW astrophysics to become a primary tool in understanding compact objects and their formation pathways. 

The basic structure of current CBC detection algorithms is based on matched-filtering, whereby candidates are identified by comparing the data against sets of pre-calculated templates, providing a discrete sampling of the parameter space, specifically by the mass of each binary component, and their dimensionless spins \citep{Wainstein_1962,Abbott_2020b}. The matched-filter correlation provides a time-series of signal-to-noise (SNR) values, and high SNR values trigger further analysis. Events with consistent parameters across detectors are assigned a ranking statistic and are subject to further analysis, including full parameter estimation of the other parameters (orientation, sky location, etc.) \citep{Abbott_2023b}.

Machine learning (ML) presents an opportunity for improving the speed, and possibly even accuracy, of GW detections and parameter estimation, enabling more rapid follow-up of any electromagnetic counterparts of future multi-messenger events \citep{Krastev_2021,Qiu_2023}. Since the first application of ML to GW detections, starting in 2018 \citep{George_2018a}, there have been several efforts to explore and enhance ML methods for the detection of CBCs. For example, \citet{George_2018a,Gabbard_2018} showed that deep learning (DL) with convolutional neural networks (CNNs) could detect {\it simulated} BBH mergers in time-series data with Gaussian noise and reach accuracies comparable to matched-filtering, as well as being more computationally efficient; this work was extended shortly thereafter using real LIGO data to detect both simulated and real BBH mergers, with similar sensitivities to matched-filtering \citep{George_2018b}. \citet{Krastev_2020} was the first to use ML to search for simulated GW signals of BNS mergers and demonstrated that a CNN could distinguish them from simulated LIGO noise and BBH mergers. In subsequent work, \citet{Krastev_2021} and \citet{Qiu_2023} extended the CNN framework to detect and estimate the parameters of GWs from BNS and NSBH mergers, respectively, using real LIGO noise.

However, these early searches only evaluated ML models on time periods with known GW signals so they could not properly account for false positives. Recently, direct ML searches for BBH signals in large sections of LIGO data have been performed, allowing for proper calculation of false alarm rates. In particular, \citet{Koloniari_2025} used AresGW and \citet{Marx_2025} used AFrame to search through O3 data for BBH mergers, detecting 40 and 38 of the 65 cataloged BBH mergers in O3 with two detectors of data.

In this paper we present an updated ML algorithm for detecting BBH mergers (extending our work in \citealt{Krastev_2020,Krastev_2021,Qiu_2023}), and carry out a full search for BBH mergers in all LIGO O1--O3 data. The paper is organized as follows.  In \S\ref{sec:methods} we describe our methods for generating training data and training time-series and time-frequency neural network models for BBH detection. We present the results of our search on both real LIGO two-detector data, and on time-shifted data (to calculate false alarm rates) in \S\ref{sec:results}. Finally, we discuss the performance of our ML pipeline, the detected and missed events, and false positives in \S\ref{sec:discussion}.

\section{Methods}
\label{sec:methods}

The overall structure of our process follows our previous work in \cite{Krastev_2020,Krastev_2021,Qiu_2023}.  We first generate a large template dataset of real LIGO noise with injected BBH signals. We then train a CNN on these data to classify between the signal and noise templates. 

Our previous papers evaluated the neural network's ability to correctly classify individual events using only the data centered around known detection times \cite{Krastev_2020,Krastev_2021,Qiu_2023}. The events were considered ``detected'' if the model assigned a probability of $p>0.5$ of the correct category. However, here we perform a ``blind'' search of the entire LIGO data stream from O1--O3. This allows us to properly take into account false positives and to calculate false alarm rates. We consider events to be ``detected'' if we recover them from all of the data, with detection thresholds chosen to minimize the false alarm rate.

In addition, instead of using only a single detector data stream, here we train models to search using both LIGO L1 and H1 data streams, In addition, in this work, we train multiple models with different training datasets and model architectures, before evaluating them on the real LIGO data to search for BBH events and correlate them with the Gravitational-wave Transient Catalog (GWTC). Finally, we also train on both time-series and time-frequency data, unlike in previous work.

\subsection{Dataset Construction}
\label{sec:dataset_construction}

We obtain LIGO data from the LIGO and Virgo Gravitational Wave Open Science Center (GWOSC) \citep{Abbott_2021a, Abbott_2023a}. Specifically, for the training data we use about $22$ days of continuous O2 data from the LIGO Livingston (L1) and Hanford (H1) detectors sampled at 4096 Hz, in which both detectors are operational and no known GW events or hardware injections are present.  To simulate BBH signals, we use the LIGO Algorithm Library Suite (LALSuite) \citep{lalsuite} to generate waveforms through the PyCBC Python library \citep{Nitz_2024}. In particular, we use the SEOBNRv4 time domain approximant \citep{Bohe_2017}.

\begin{figure}[t!]
    \centering
    \includegraphics[width=0.48\textwidth]{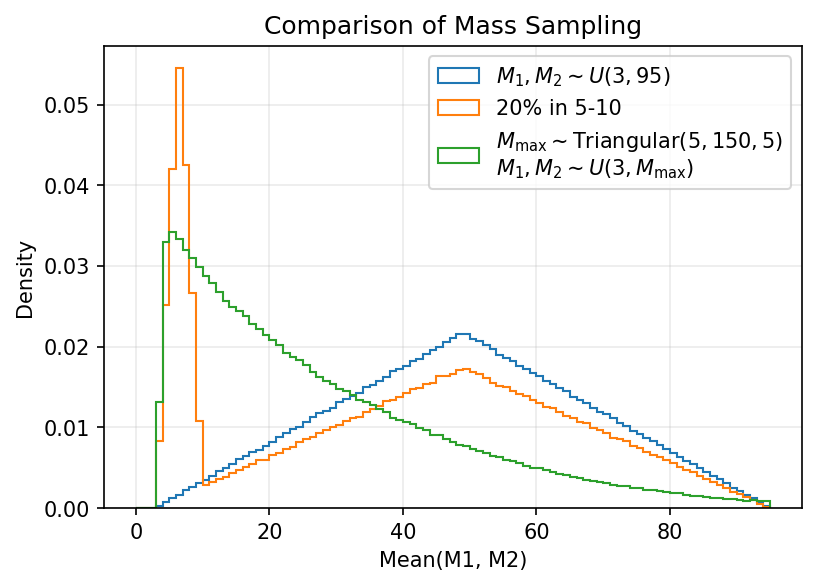}
    \caption{Comparison of the various BBH mean mass distributions of the training data for models we have tested. We first show sampling each mass independently from a uniform distribution. Second, is doing the same except forcing 20\% of those to be in the range (5-10M$_\odot$). Third, is the mass distribution we use in the datasets for the models shown in this paper, as described in the text.}
    \label{fig:mass_dist}
\end{figure}

We inject BBH waveforms in half of the original noise templates. To guard against potential lower performance at lower mass, we want a sufficient number of templates with two low-mass black holes; if we sample each component mass individually from a uniform distribution, few events will have both black holes at low mass. Therefore, we choose to over-sample low-mass BBHs in the following way: first we sample the maximum mass from a triangular distribution spanning $5-150$ M$_\odot$ with a peak at $5$ M$_\odot$ ($\textrm{Max}=\textrm{Triangular}(5,150,5)$ M$_\odot$). We then sample each mass from a uniform distribution as $M_1,M_2=\textrm{Unif}(3,\textrm{Max})$. We also impose a maximum mass ratio $M_1/M_2 \leq 10$; see Figure~\ref{fig:mass_dist}.

\begin{figure*}[t!]
    \centering
    \includegraphics[width=0.95\textwidth]{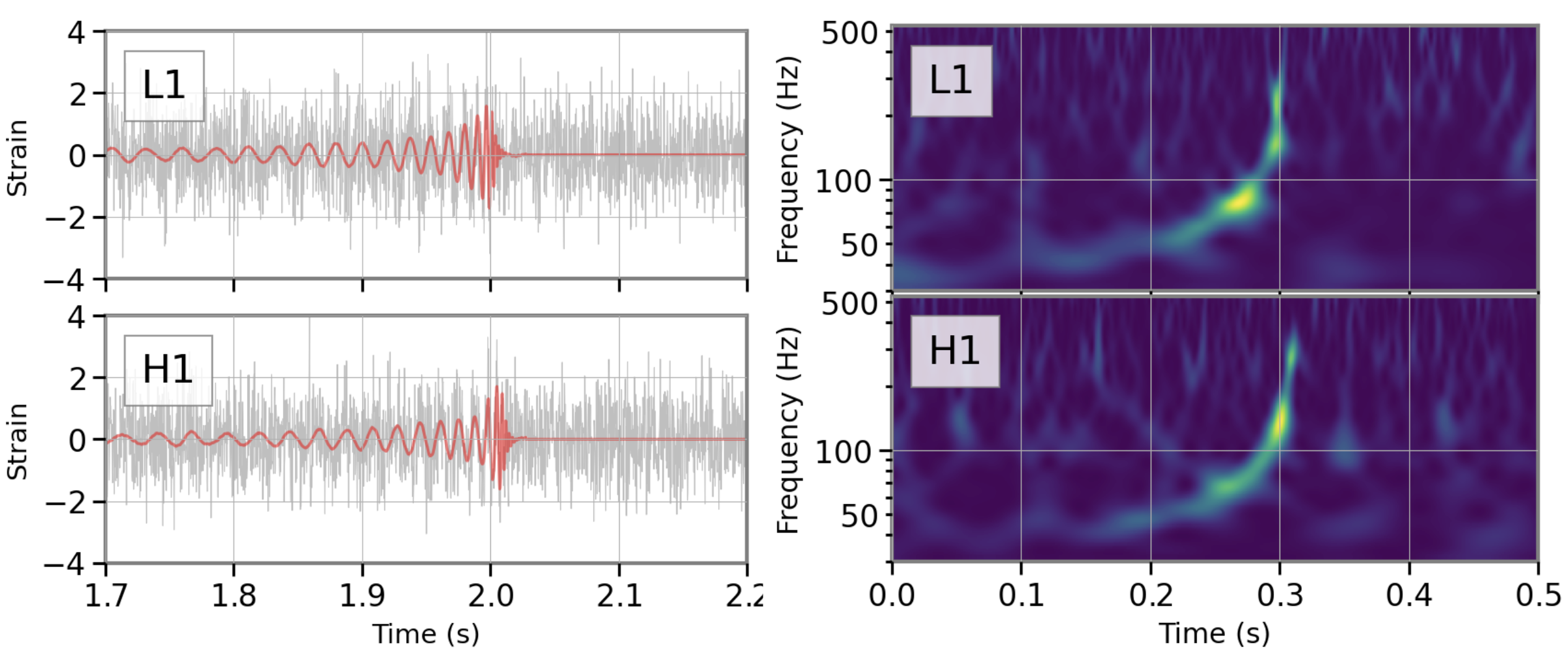}
    \caption{{\it Left:} Example of a BBH template injection in LIGO noise from our training data. The template has $M_1=25$ M$_\odot$, $M_2=20$ M$_\odot$, and a network ${\rm SNR}=10$. The red lines are the BBH waveforms alone, and the gray lines are the combined BBH waveform and noise. {\it Right:} The time-frequency data corresponding to the same template. We perform the Q-transform of a 0.5 s window around the signal peak of the time-series data to convert it to a $300x300$ time-frequency image. The signal chirp and merger are clearly seen.}
    \label{fig:data_example}
\end{figure*}

For all waveforms, we assume zero spin and zero eccentricity. We sample the waveforms at 4096 Hz for 4 s. After projecting the waveforms to each LIGO detector, we shift them in time to position the peak of the signal randomly in the window of $3.7-3.9$ s within the 4 s template, or $3.79-3.81$ s for high-resolution models (see \S\ref{sec:high_res_models}). We whiten the noise data and the simulated signals separately with the power spectral density (PSD) computed directly from the raw GW strain data using Welch’s method \citep{Welch_1967}. We scale the amplitude of the waveforms to reach a certain sampled SNR level; sampling SNR is equivalent to sampling distance, but we find it effective to draw SNR from a uniform distribution: $\textrm{SNR}=\textrm{Unif}(7,20)$.  An example of a generated BBH template is shown in Figure~\ref{fig:data_example}.

We generate 480,000 templates for training for time-series data, and 160,000 for time-frequency data (see Section \ref{sec:time-frequency}). We also use 20,000 templates for validation. All of the training and validation templates use fully unique segments of real LIGO noise. Each dataset is 1/2 noise with no waveform and 1/2 noise + BBH signal.

\subsection{One- and two-detector models}
For all of the neural network models we train, we train and evaluate them on two detectors of data. However, we have two ways of actually implementing this. We can have one neural network evaluated on two-detector data (``two-detector models'', or two neural networks each evaluated on one detector, with coincidence enforced between them (``one-detector models''). This is done by taking the minimum of the BBH probability output of each neural network (i.e. a detection must be recorded in each detector at the same time).

These two approaches are combined because the two-detector models are able to better incorporate data from both detectors simultaneously, and can reject signals that are inconsistent in the two detectors. However, the one-detector models are more strict on enforcing coincidence, which is helpful for avoiding false positives.

\subsection{Time-Frequency Templates}\label{sec:time-frequency}

While using the 1D time series template data is useful, we find that it is beneficial to combine this with 2D time-frequency data, computed with the Q-transform available with the GWpy package \citep{gwpy}.  These templates are generated in the exact way as the time-series data, but with the Q-transform computed from 30 to 330 Hz, using the last 0.5 s of the template, and generating a $300x300$ array. This can cut off the early part of some longer-duration waveforms, but we choose this size as a balance between duration, resolution, and data size. An example of this process is shown in Figure \ref{fig:data_example}, which shows a 0.5 s window around the signal peak in the time-series data with its corresponding Q-transform.

\begin{figure*}[t!]
    \centering
    \includegraphics[width=0.95\linewidth]{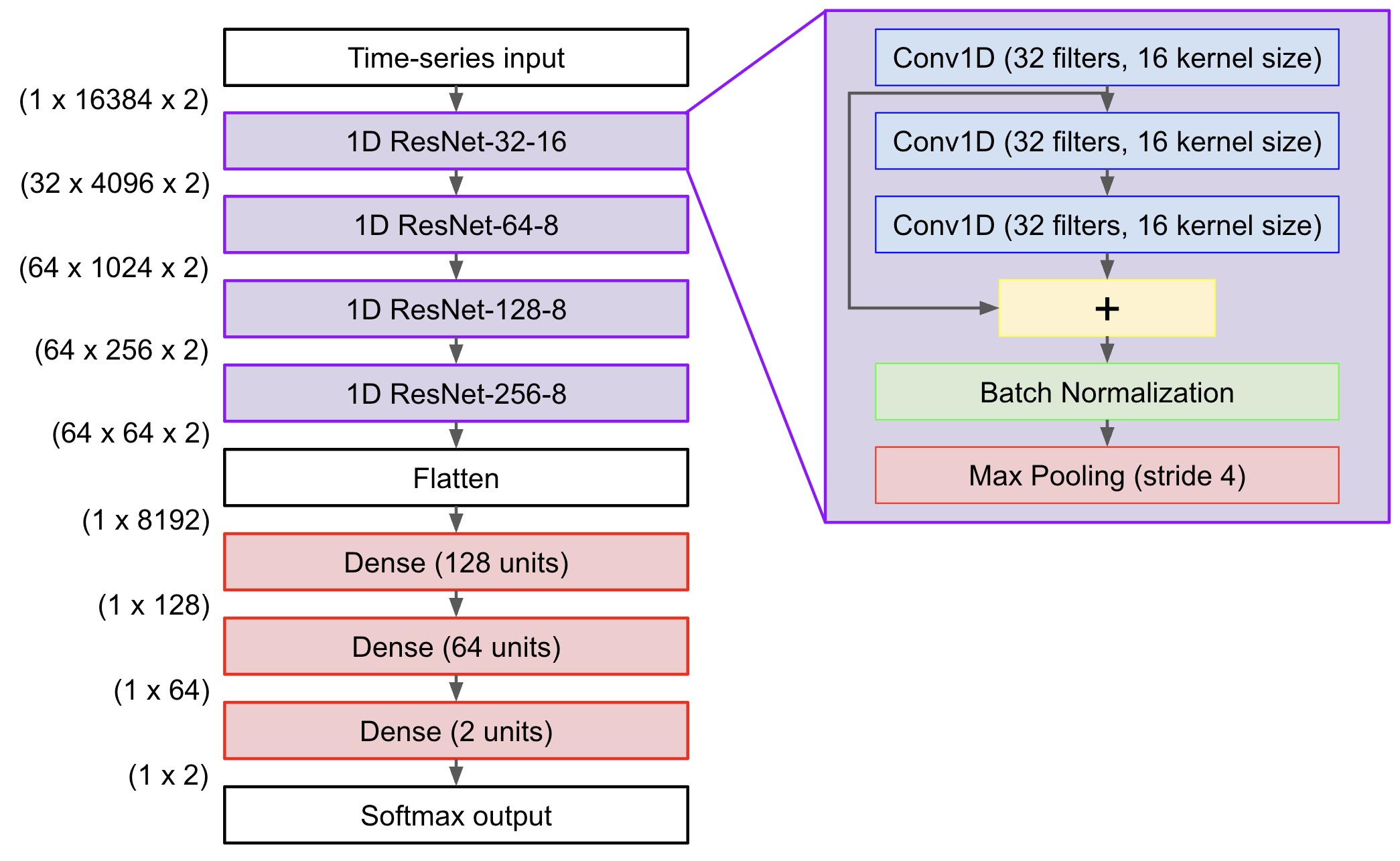}
    \caption{Schematic diagram of our residual neural network architecture for two-detector time-series ResNet models. This model contains 3,862,756 parameters.}
    \label{fig:architecture_1d}
\end{figure*}

\begin{figure*}[t!]
    \centering
    \includegraphics[width=0.95\linewidth]{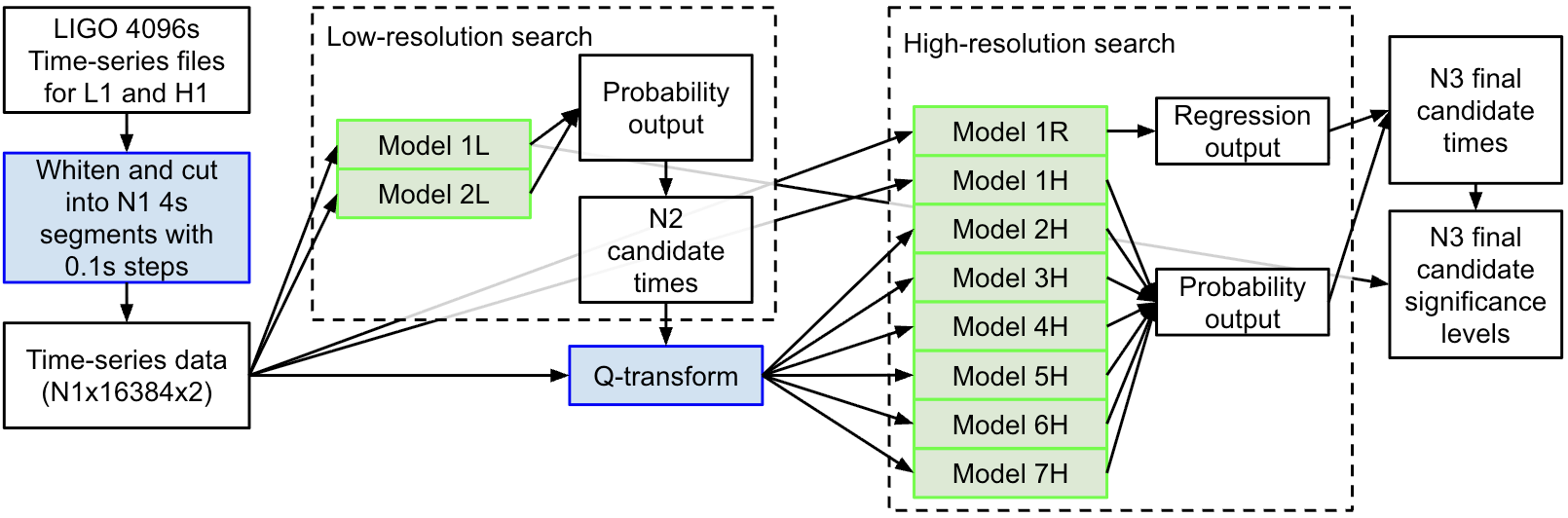}
    \caption{Flow-chart of our GW search process, starting from the initial LIGO files, data whitening and splitting into 4-s segments, evaluation through the low- and high-resolution time-series and time-frequency models, and  generation of a final list of candidates.}
    \label{fig:search_flowchart}
\end{figure*}

\subsection{High-Resolution Models}
\label{sec:high_res_models}

Our BBH search in the O1--O3 data is first performed at lower time resolution, using steps of 0.1 s. These models select an initial list of candidates, which are then evaluated with higher resolution models, using steps of 0.01 s. This approach greatly speeds up the search because not every model has to be evaluated on all of the available data. Since the models trained on time-series data are much faster in evaluation, only those models are used in the initial low-resolution step, while all time-frequency models are used in the high-resolution step on event candidates. As explained in \S\ref{sec:dataset_construction}, the peak of the generated waveform is placed randomly in a window near the end of a 4-s template, $3.7-3.9$ s for the low-resolution models or $3.79-3.81$ s for the high-resolution models. These time windows are designed to ensure that the signal peak will be captured in 2 time steps.

\subsection{Generated False Positives}\label{sec:generated_fps}

To help train the neural network to identify and avoid various types of false positives, we add four classes of false positives as templates labeled as noise in the training data. These false positives are only added to the two-detector models, as one-detector models do not view both data streams simultaneously. The injected false positives include the following cases:
\begin{enumerate}
    \item \textbf{Single-detector signals}: We  inject the BBH waveform into only one detector, so that the other detector shows only noise. This is designed to teach the neural network to require a signal in both detectors.
    \item \textbf{Different parameters}: We inject independently generated BBH waveforms to each detector, with the condition that the chirp mass of the two waveforms must differ by at least 50\%. This is designed to teach the neural network to avoid cases in which each detector presents a plausible BBH signal, but the two signals are inconsistent with each other.
    \item \textbf{Time offset}: We inject the same BBH waveform to each detector, but offset in time by more than 50 ms.  This is designed to teach the neural network to avoid signals with time offsets that are too large to have been caused by a single event.
    \item \textbf{Coincident noise}: For the time-frequency data, we locate noise spikes throughout the data, and then create artificial templates by aligning those spikes of noise in the sensitive window. This is designed to teach the neural network to avoid random noise that happens to align closely in time in both detectors.
\end{enumerate}

\subsection{Chirp Mass Regression Model}

To help ensure that any candidates we identify in the search are consistent with BBH waveforms in both detectors, we train a regression model to predict the chirp mass from a single-detector signal. The data for this regression model are similar to the other generated datasets, but include no false templates, no false positives, and an SNR range of $0-20$.  By comparing the predicted chirp mass from each detector's data, we can rule out some candidates that produce substantially different values between the detectors. We do this by calculating a value we call the ``mass similarity'' that determines how similar the predicted chirp masses are:
\begin{equation}
    \mathcal{MS}=-\log\frac{\mathcal{M}_1-\mathcal{M}_2}{(\mathcal{M}_1+\mathcal{M}_2)/2}
\end{equation}
We calculate this $\mathcal{MS}$ for each candidate, and rule out the candidates with large chirp mass differences of $\mathcal{MS}<0.4$.

\subsection{Neural Network Architecture and Training}

For our neural network models, we use a structure similar to that of \cite{Qiu_2023} for all models. We use several different model architectures for time-series, using 1D layers, and time-frequency, using 2D layers (see Table \ref{tab:pipeline_models}). The ``CNN'' architecture contains 4 convolutional layers with ReLU activation, each followed by a max pooling layer, then followed by 2 dense hidden layers. The number of filters in each convolutional layer is 32, 64, 128, and 256, respectively. For the time-series models, the kernel sizes in each ResNet block are 16, 8, 8, and 8, respectively, and for the time-frequency models, the kernel sizes are all 3. The max pooling layers are size 4 for time-series and size 2 for time-frequency. The dense hidden layers have widths of 128 and 64. The final softmax output corresponds to the number of predicted classes, 2, defined as BBH and noise. 

Most of our models use a ResNet structure, which involves residual connections: the output of an earlier layer in a block is added to the output of a later layer. These residual connections are designed to make deeper neural networks easier to optimize and achieve high accuracies \citep{He_2016}. In our ResNet architecture, each of the 4 convolutional layers in our CNN architecture is replaced by a ResNet blocks of 3 convolutional layers and a batch normalization layer each, as shown in the right panel of Figure \ref{fig:architecture_1d}. For one model, we add a dropout layer with a rate of 0.5 before the dense layers. As an example, the schematic diagram of the architecture of the two-detector time-series ResNet model is shown in Figure \ref{fig:architecture_1d}. It has 3,862,756 parameters in total. Finally, we point out that the exact hyperparameters we use for our models are unlikely to change the results significantly. We have chosen the specific hyperparameters to give good performance, but a thorough hyperparameter optimization would only have a very small effect on the pipeline's results.

We build and train our neural network models using TensorFlow \citep{Tensorflow_2015}. We use the Adam optimizer \citep{Kingma_2017} with AMSgrad \citep{Reddi_2019}. We use $\alpha=0.001$ as our learning rate, $\beta_1=0.9$, $\beta_2=0.999$, $\epsilon=10^{-8}$, and a batch size of 216. We use sparse categorical cross-entropy loss and train for 200 epochs. Our final models are taken from the epoch with the lowest validation loss. We train our models using 4 NVidia A100 GPUs with a data-parallel strategy.

\section{Results}
\label{sec:results}

\subsection{Model Training Results}\label{sec:model_training_results}

\begin{table*}
\centering
    \begin{tabular}{ccccccc}
    \hline\hline
    Model & Res. & \# Det. & Data & FPs & Architecture & Threshold\\
    \hline
    1L & Low & 1 & Time &  None & ResNet & 0.54\\
    2L & Low & 2 & Time &  None & ResNet & 3.2\\
    1H & High & 2 & Time-Freq. &  None &  CNN & 2.1\\
     2H & High & 2 & Time-Freq. &  None & ResNet & 1.9\\
     3H & High & 2 & Time-Freq. &  1--3 &  CNN & 3.1\\
     4H & High & 2 & Time-Freq. &  1--3 & ResNet & 0.92\\
     5H & High & 2 & Time-Freq. &  1--3 & ResNet w/dropout & 0.84\\
     6H & High & 2 & Time & 1--4 & ResNet & 12.5\\
     7H & High & 2 & Time & 1--4$^*$ & ResNet & 8.0\\
     1R & High & 1 & Time & None &  CNN (Regression) & $0.4$ \\
    \hline\hline
    \end{tabular}
    \caption{Models used in the search pipeline and their characteristics. The ``\# Detectors'' column refers to whether the model is a one-detector or two-detector model. The FPs column is which types of generated false positives from Section \ref{sec:generated_fps} are added to noise templates. For models with FPs, 50\% of noise templates have added FPs, except Model~7H, for which 100\% of noise templates have added FPs.} 
    \label{tab:pipeline_models}
\end{table*}

We perform tests of many different models, which differed in the training data setup, the false positives added, the neural network architecture, and the training details. After initial testing each model, we arrive at a final list of models that are most effective in finding BBH events; see Table \ref{tab:pipeline_models}.

To determine if a particular event is ``detected'' by a model, we do not simply use a 50\% probability threshold. This is because our training data uses a 50\% ratio of GW events, while in the actual data real BBH events are exceedingly rare. As a result, the model's predicted probabilities for true events are often very close to unity, so we define the significance as $S=-\log(1-p)$; e.g., $p=99\%$ translates to $S=2$, $p=99.9\%$ to $S=3$, etc.

Furthermore, because the training data of each model varies, the probability assigned to a single event by different models might vary greatly.  Therefore, we use a decision threshold for each model based on its performance in the overall GW search (see \S\ref{sec:search_results}), shown in the last column in Table \ref{tab:pipeline_models}.

\subsection{GW Search Pipeline}
\label{sec:gw_search_pipeline}

In our GW search, we evaluate all available two-detector data from LIGO's observing runs O1, O2, and O3, with the goal of finding as many of the true (i.e., cataloged) BBH events as possible with as few false positives (FPs) as possible. 

To start, we whiten the available data to match how we set up the training data, and exclude times with a data quality or injection flag (these flags are provided at 1 s intervals).
The first step in our search pipeline is evaluating the low-resolution models (Models 1L and 2L) on the data in 0.1 s time-steps. We identify candidate BBH events as those detected by both of our low-resolution models.
We then select the 0.2 s region around each candidate and take the Q-transform to prepare the time-frequency data.
We evaluate the high-resolution models in 0.01 s time-steps for both the time-series and time-frequency models. Candidates that are also detected by all of the high-resolution models are selected as the final list of events detected by our pipeline.

A flow-chart of our overall search process is shown in Figure~\ref{fig:search_flowchart}. Using the conventions in the flow-chart, a typical 4096-s LIGO file will generate $N_1\approx40,000$ segments. Over all observing runs, we find $N_2=416$ initial candidates from the low-resolution search. Then, the high-resolution search narrows this down to $N_3=114$ final candidates (if all high-resolution models are used) of which $57$ are true positives and $57$ are false positives. In this paper, we define ``true positives'' as BBH events in the GWTC-3 catalog with two detectors of data that our pipeline detects, and ``false positives'' as those that our pipeline does not detect.

Finally, after the final candidates are selected, we choose the output of the one-detector Model 1L as the final significance level. This is because after testing the output significances of each of the models, and many combinations of them (including the sum/median of all the significance levels or the product of the probabilities), we find that the output of Model 1L best separates true and false positives.

\subsection{GW Search Results}\label{sec:search_results}

The initial low-resolution search is the step that must be the most selective, as these models are evaluated on all of the data, so even a very low false alarm rate will give many times more false positives than true positives.
Therefore, we must choose a tradeoff of identifying the greatest fraction of the true positives (BBH events identified by LVC: LIGO-Virgo Collaboration) as possible while minimizing the number of false positives.

There are 75 BBH events identified by the LVC with two detectors of data available in O1, O2, and O3.
At our chosen thresholds we identify 57 of these events. We examine the events we miss further in \S\ref{sec:missed_tps}, but most have low SNR in one detector, a low network SNR, or a low chirp mass. Events are only missed during the initial low-resolution search step, which is quite selective to minimize the false alarm rate. We note that one of the 57 detected events, GW200129\_065458, was originally missed in our search due to a data quality flag. However, it is a very significant event, and would be detected easily without that flag, so we decide to include it as detected.

In addition to these true positives, our model pipeline gives 57 false positives, with a significance at least as high as the least significant true positive, when all high-resolution models are included. We examine some of these false positives in \S\ref{sec:fps}.

\begin{figure*}
    \centering
    \includegraphics[width=\textwidth]{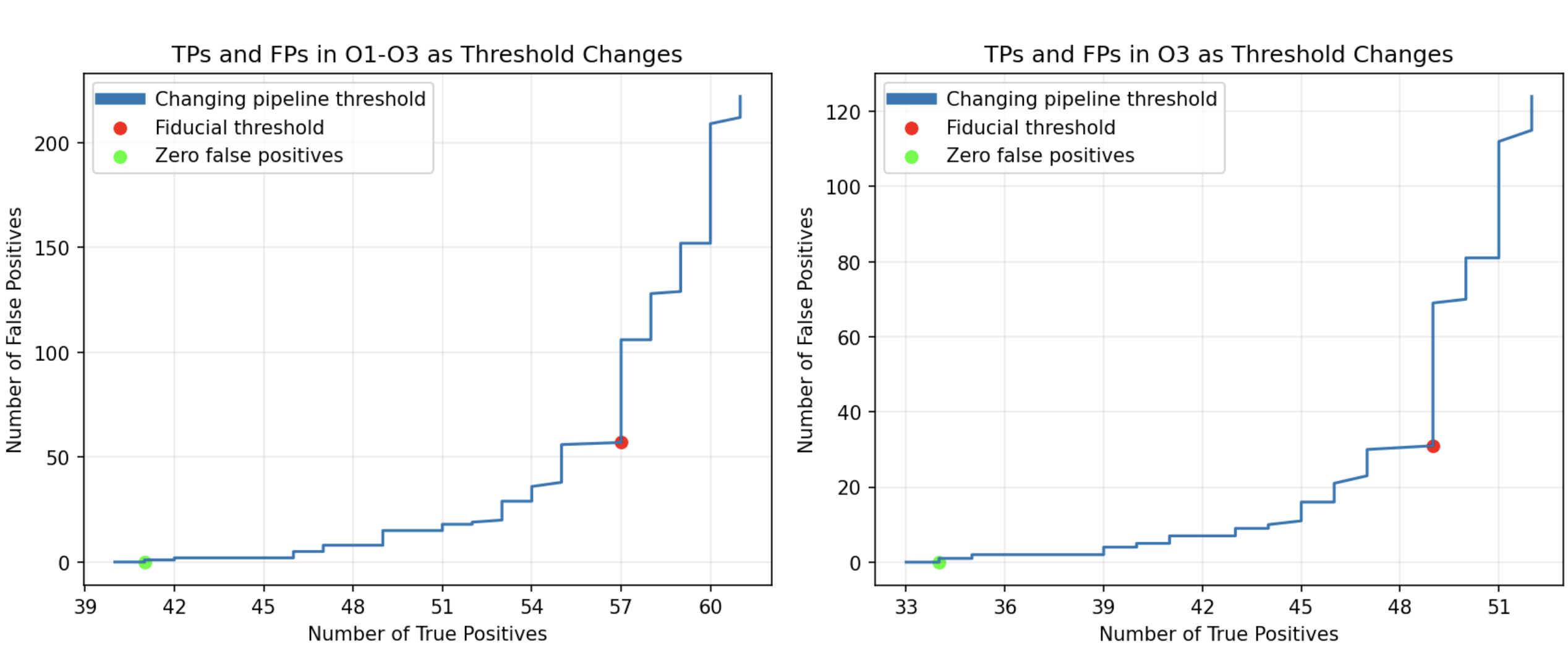}
    \caption{The number of false positives and true positives for our model pipeline as the final threshold changes, in all of O1-O3, and O3 specifically.}
    \label{fig:threshold_changes}
\end{figure*}

A plot of the number of false positives and true positives for the pipeline as threshold changes is shown in Figure \ref{fig:threshold_changes}. This figure shows that depending on the desired search results, the pipeline can be flexibly adapted by changing the final pipeline threshold (the Model 1L threshold). For this paper, we choose a threshold of 0.54, which gives 57 FPs and 57 TPs as a balance between detected most of the BBH events without too many FPs. However, if greater purity is desired, for example, the threshold can be increased to give 0 FPs, at the cost of only detecting 41 TPs out of 75 total BBH events in O1--O3 with 2 detectors of data.

\subsection{Time-Shifting and FARs}
\label{sec:time_shifting}

To determine the false alarm rate (FAR) of our pipeline, and to calibrate the FAR relative to the model significance, we time-shift the LIGO data 100 times, and evaluate the entire model pipeline on each time-shift.  Specifically, we shift the H1 data stream in consecutive steps of 2 s relative to the L1 data stream.  Since in the time-shifted data all events are false positives, we can empirically determine FAR as a function of model significance.

\begin{figure*}[t!]
    \centering
    \includegraphics[width=0.95\textwidth]{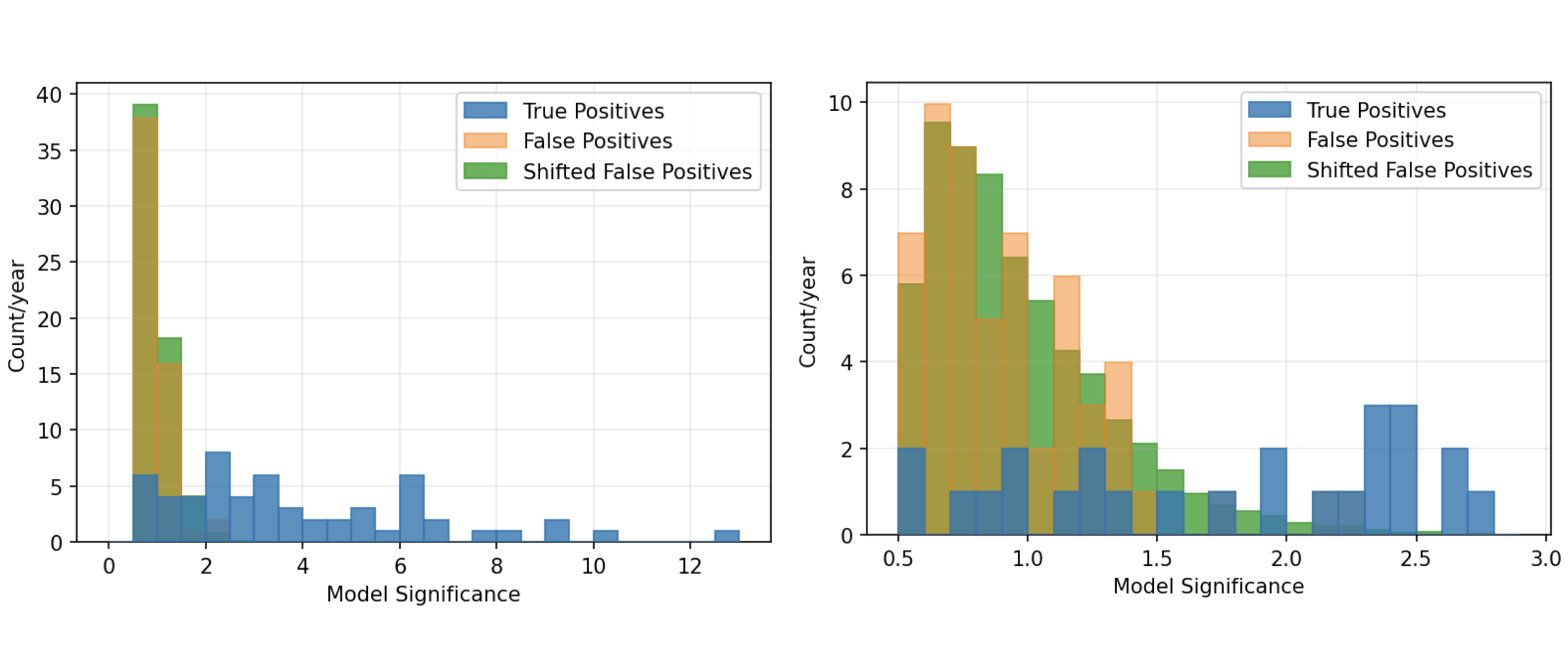}
    \caption{Histograms of true positives (blue) and false positives (orange) from the O1--O3 search, as well as false positives from the time-shifting analysis (green).}
    \label{fig:far_hist}
\end{figure*}
\begin{figure}[t!]
    \centering
    \includegraphics[width=0.48\textwidth]{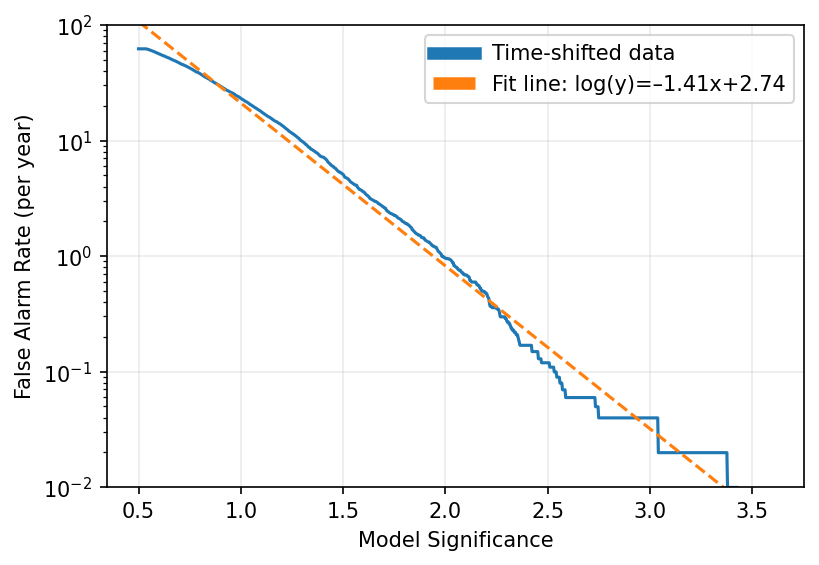}
    \caption{Cumulative histogram of the false alarm rate (FAR) as a function of model significance from the ensemble of 100 time-shifted data sets (blue).  The orange line shows a simple power law fit (linear in log-space).}
    \label{fig:far_fit}
\end{figure}

In this time-shifted search, totaling about $100$ years of data, we find $6,273$ false positives above a model significance of $S=0.54$ (the threshold for Model 1L); the highest model significance we find in the distribution is $S=3.43$. We show a histogram of the distribution of model significance of these time-shifted false positives compared to the real-data true and false positives in Figure~\ref{fig:far_hist}.  The cumulative distribution is shown in Figure~\ref{fig:far_fit}. Of these 6273 false positives, 31 have one detector of a true event's data present. This is a small fraction ($<0.5\%$), and these false positives are not greatly more significant than average, so they do not impact the total significantly (The maximum significance reached is $S=1.78$). To estimate the FAR for all true events, including those more significant than any of the false positives in our time-shifted distribution, we extrapolate the observed FAR distribution using a power law fit (linear in log-space) (Figure \ref{fig:far_fit}).

\section{Discussion}
\label{sec:discussion}

\subsection{Identified and Missed BBH Events}\label{sec:missed_tps}

Our search pipeline identifies 57 of the 75 total BBH events in the GWTC-3 catalog with two detectors of data in O1--O3.\footnote{In addition to these 75 events, there is an additional event, GW170608, published in GWTC-3 which did not have two detectors of data in the original data stream, but LVC has still published data from both detectors. This is because H1 was being tested while only L1 was in normal operation, but data for both detectors was still recorded. Our model detects GW170608 at a $6.10$ significance (implying a very low FAR of $\sim1/350,000$ years).} These include events over a wide range of network SNRs (4.5--26.8), component masses (5.1--105.5$M_\odot$), and other properties. However, as discussed below, the fraction correctly identified decreases for low-SNR and low-mass events.

\begin{figure}[t!]
    \centering
    \includegraphics[width=0.48\textwidth]{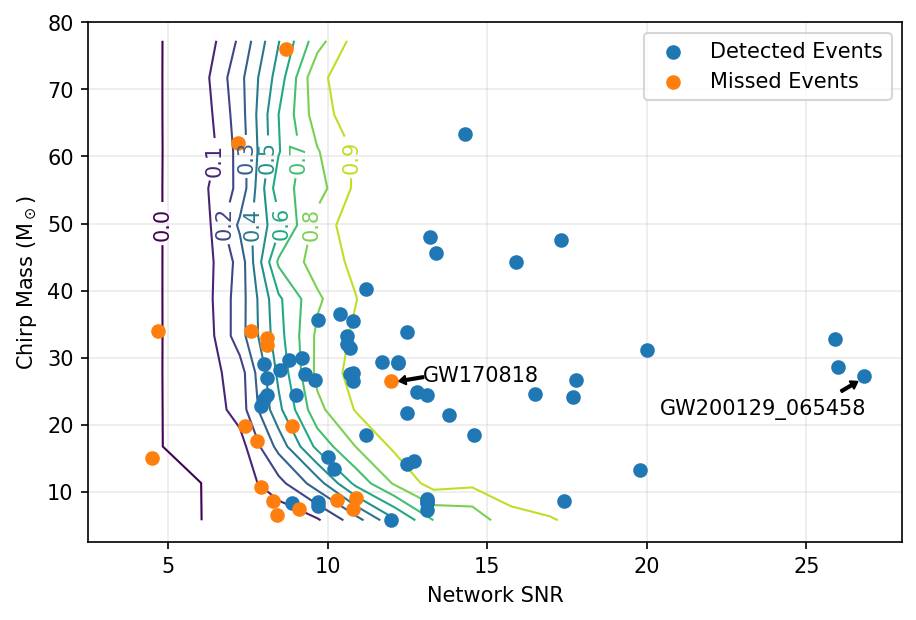}
    \caption{Chirp mass versus network SNR for BBH events from GWTC-3 with data in both L1 and H1. Blue points indicate events recovered by our search pipeline, while orange points are the missed events. The contours show the accuracy of the search pipeline on validation data as a function of chirp mass and SNR.  We find that missed events are mainly at low SNR and low chirp mass, where our pipeline performance is degraded. The real recovery rate is in good agreement with the expected performance based on our validation data.}
    \label{fig:missed_vs_detected}
\end{figure}

Our search pipeline does not recover 18 of the 75 BBH events in GWTC-3. The most common cause of a BBH event being missed is low network SNR of $\lesssim 10$, and most missed events are also concentrated at low chirp mass of $\mathcal{M}\lesssim 10$ M$_\odot$; see Figure~\ref{fig:missed_vs_detected}.  All 18 events are missed by the initial low-resolution search models, i.e., by at least one of models 1L and 2L. Since these are one- and two-detector models, and all models must detect the event, this can occur due to a low SNR in one detector, or a low network SNR.  In this vein, GW170818 stands out as a missed event with a fairly high network SNR, but in this case nearly all of the SNR is in L1, with a weak signal in H1. Thus, this event is missed in the one-detector model L1, which requires the event to be detected in both detectors.

In Figure~\ref{fig:missed_vs_detected} we also show the performance of the search pipeline evaluated on a set of $100,000$ validation templates, generated with a uniform distribution of ${\rm SNR}=3-20$ and a uniform distribution in mean mass from 3--95$M_\odot$. The resulting contours mark the detected fraction, and show that at $\mathcal{M}\gtrsim 20$ M$_\odot$ the detection fraction depends only on network SNR, with $50\%$ efficiency at ${\rm SNR}\gtrsim 8$; at lower chirp masses the pipeline performance degrades so that by $\mathcal{M}\approx 10$ M$_\odot$ the required SNR for $50\%$ efficiency is $\approx 12$.  The contours also roughly match the trend seen in the real data.

Although our low-resolution search missed 18 events, we can still evaluate them with the high-resolution models by targeting the specific times of these events.  Of the 18, 2 are detected by all of the high-resolution models.

\begin{figure}[t!]
    \centering
    \includegraphics[width=0.48\textwidth]{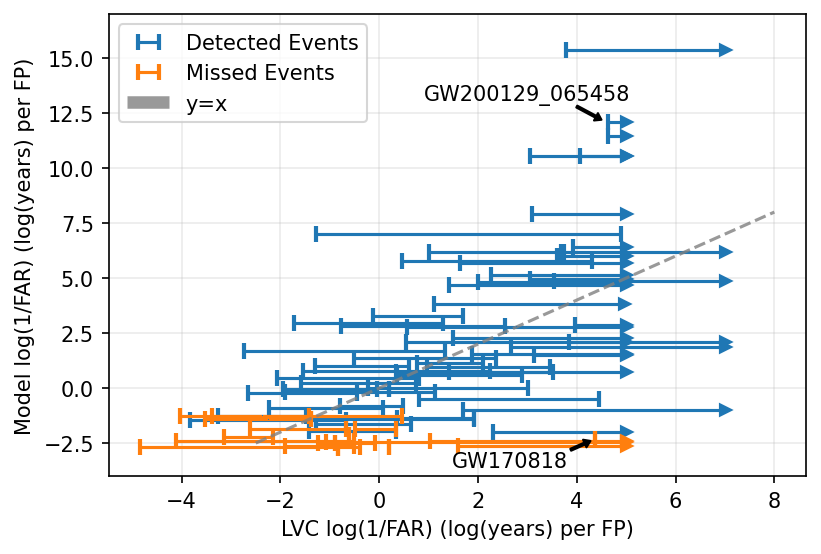}
    \caption{The FAR from our analysis (extrapolated from the fit shown in Figure~\ref{fig:far_fit}) versus the FAR provided in GWTC-3 for all BBH events; events recovered by our pipeline are shown in blue, while missed events are shown in orange.  For the GWTC-3 FAR values we show the range from the various pipelines listed in the catalog.  We find an overall correlation between the FAR values of the two independent analyses.}
    \label{fig:far_comparison}
\end{figure}

With the fit of the false alarm rate calculated in Section \ref{sec:time_shifting}, we can now estimate the FAR on each true event. We compare the FAR we calculate to the ones calculated for GWTC-3 in Figure \ref{fig:far_comparison}. If one considers the lowest FAR of the LVC pipelines, more of our estimates give higher FARs, so the best combination of the LVC pipelines generally gives lower False Alarm Rates, especially for the less significant events. But if one considers the average of the pipelines used by LVC, our estimated FAR values are more similar, and our extrapolation to very low FAR values (the upper part of the graph) are lower than LVC's (these extrapolated values could be underestimated compared to LVC, which usually uses a more conservative limit of $10^{-5}/$yr). 

\subsection{False Positives}
\label{sec:fps}

\begin{figure*}
    \centering
    \includegraphics[width=0.95\textwidth]{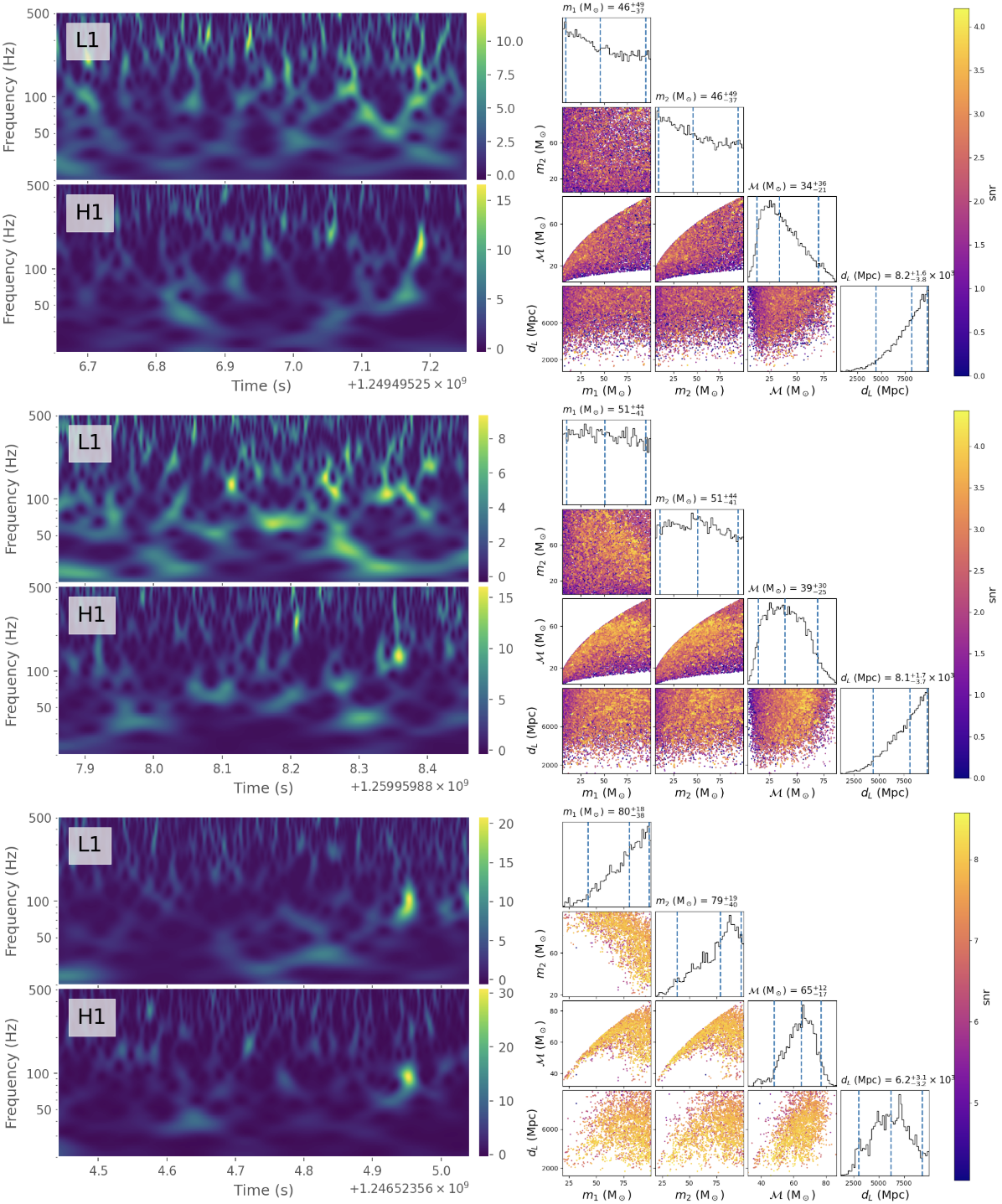}
    \caption{The three most significant false positives from our BBH search. The left side shows time-frequency data for each detector, and the right shows an example of performing parameter inference on the false positive.}
    \label{fig:fp_examples}
\end{figure*}

As mentioned in \S\ref{sec:search_results}, our full search pipeline leads to 57 false positives with a significance at least as high as the least significant true positive. We label the candidates not identified in GWTC-3 as FPs, but note that some may be real events, such as with a low SNR (too low to confirm as a true event), or otherwise not distinguishable as a true event from false positives. We show the three most significant false positives in Figure~\ref{fig:fp_examples}. They have significances of 2.28 ($\textrm{FAR}=0.30/$yr), 2.16 ($\textrm{FAR}=0.56/$yr), and 1.75 ($\textrm{FAR}=2.24/$yr), respectively.

The most common cause of false positives appears to be simply when transient noise happens to align closely in both detectors. We have trained some of the neural network models to avoid this, but it is very hard to prevent entirely. However, many of these FPs can be ruled out by parameter inference. An example of this parameter inference for the three most significant false positives, is shown in Figure \ref{fig:fp_examples}.

The two most significant events in Figure~\ref{fig:fp_examples} both show very poor convergence of the parameter inference. This indicates that parameter inference was not able to find a realistic solution to both detectors of data.
The third FP has inference that does not converge well, but it is better than the other FPs, and not much worse than real low-SNR events. However, it can be excluded because it is such a short event that it must be high mass, but real high mass events typically start at a lower frequency, usually at most 30--40 Hz, and are still not quite so short. 

Very fast events like the third FP sometimes can't be excluded just from the parameter inference because there is less data to use to rule out the event. However, these events can still usually be excluded by a combination of human inspection and more advanced false positive checks.

\subsection{Model Complexity Tradeoffs}

\begin{figure}[t!]
    \centering
    \includegraphics[width=0.48\textwidth]{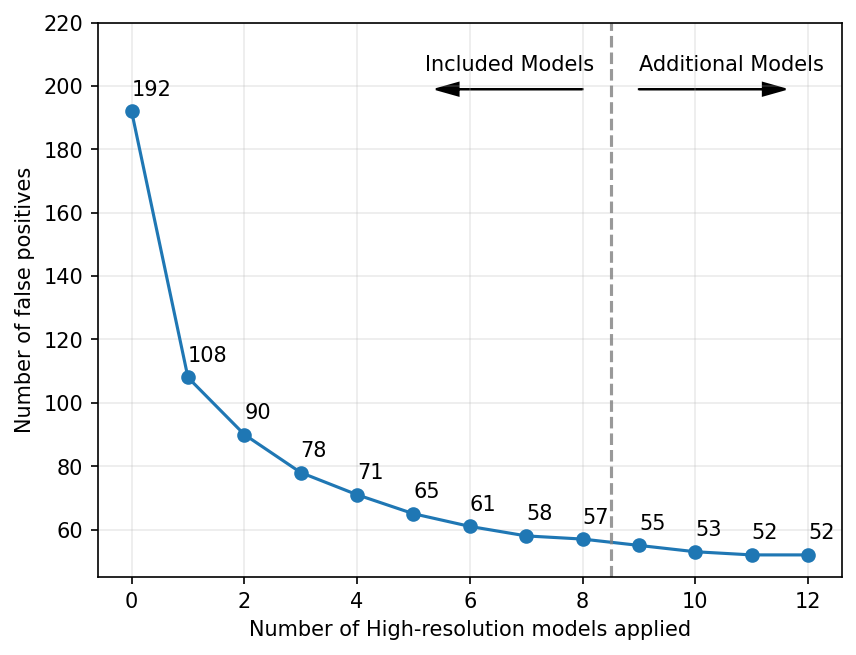}
    \caption{The number of false positives as a function of the number of neural network models in the high-resolution search pipeline. When each model is added, its threshold (from Table \ref{tab:pipeline_models}) is low enough to not exclude any true positives, so all points detect the 57 TPs of our final pipeline. First the 8 high-resolution models we include in this paper are added, and then we include the next four best models, added sequentially based on how many false positives they remove.}
    \label{fig:fp_staircase}
\end{figure}

Our search pipeline uses 10 total models in an ensemble, each with its own threshold (see \S\ref{sec:model_training_results} and Table \ref{tab:pipeline_models}).  The number of models used impacts the balance of true and false positives; e.g., a simplified version with fewer models in the ensemble might be preferable for flexibility, but leads to a higher number of false positives. To assess the impact of our choice of models number, we varied the number of sequential high-resolution models and determined the number of resulting false positives; see Figure~\ref{fig:fp_staircase}. We find that after the first 7 models or so used, additional models added only provide a very small benefit, sometimes only removing a single FP. We settle on 8 high-resolution models as the point where adding more gives little benefit for increasing the complexity of the pipeline.
\footnote{The 8th high-resolution included is the chirp mass regression model, which only gives a small improvement in the pipeline. However, we include the model because it does exclude more FPs in the time-shifted search, decreasing our False Alarm Rates, and to serve as a proof of concept for incorporating parameter estimation into our pipeline.} The best number and combination of the models in our pipeline could be changed to adapt to different applications or datasets.



\begin{figure*}
    \centering
    \includegraphics[width=0.95\textwidth]{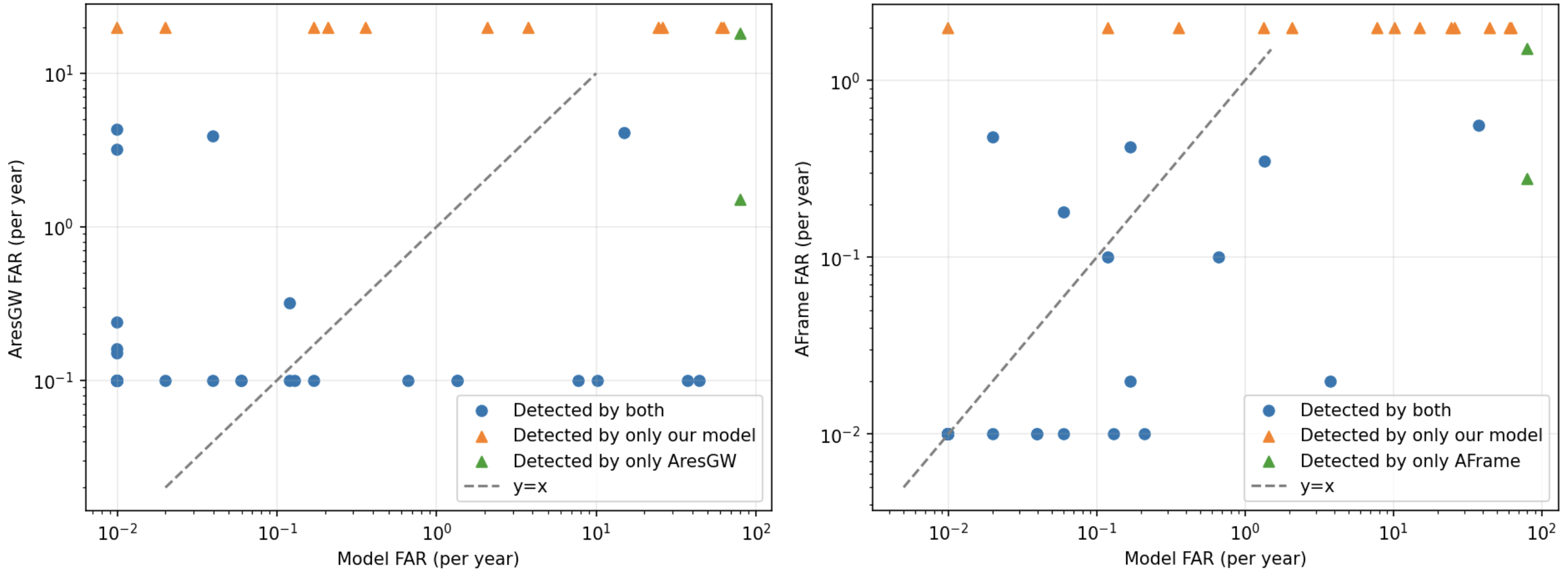}
    \caption{The left and right plots show comparisons of the False Alarm Rates (FAR) calculated by our search, compared with AresGW and AFrame, respectively. For consistency, the FARs from all three searches are limited by time-shifted searches ($0.01/$yr for our search and AFrame, which used 100 years of time-shifted background, and $0.1/$yr for AresGW, which used 10 years). For the points detected by only one model, of course there is only one FAR calculated, so they are shown near the axis. For AresGW, there are 16 points at (0.01, 0.1), and for AFrame, there are 21 points at (0.01, 0.01).}
    \label{fig:search_comparison}
\end{figure*}

\subsection{Comparison to Other Searches}

Now, we will perform a brief comparison of the results of our search to the two most comparable ML-based searches that have been done, with the algorithms AresGW \citep{Koloniari_2025}, and AFrame \citep{Marx_2025}. Both AresGW and AFrame only searched for BBH events in O3, while we included O1 and O2. With $p_{\textrm{astro}}>0.5$, AresGW detects 40 of the 65 total BBH events in GWTC-3 with two detectors of data and no data quality issues, while AFrame detects 38 of them. Our search detects 49 of the events, although some of those are detected with fairly high False Alarm Rates. 

A comparison of the FARs calculated by each search is shown in Figure \ref{fig:search_comparison}. This figure shows that our pipeline achieves lower FAR values than AresGW for most events, and can detect 9 more events in total. AFrame is able to achieve very low FAR values, with AFrame and our pipeline achieving $\textrm{FAR}=0.01/$yr for 21 BBH events. AFrame's FARs are lower than our pipeline's results for most of the other events we both detect, but we are able to detect 11 more BBH events in total in O3 than AFrame. Thus, while AFrame is usually as confident or more confident on the events both pipelines detect, our pipeline detects more in total.

Comparing an equal number of detected events, we see that AresGW reaches 38 events detected at a maximum FAR of $4.3/$yr.  AFrame reaches 38 events detected at a maximum FAR of $1.52/$yr, while our pipeline reaches 38 events detected at a maximum FAR of $1.34/$yr.

Comparing these three search algorithms, there are minor differences in model architecture, training data, and search procedure. Other differences include more consistency checks used by AresGW, and that AresGW and AFrame use more sophisticated calculations of $p_{astro}$. The main advantage of our pipeline is most likely that it takes advantage of multiple neural network models in an ensemble to increase sensitivity, including time-frequency data in the high-resolution search. While the searches are all comparable in performance, our pipeline is able to detect the greatest number of BBH events.

\section{Summary and conclusions}

In this work, we have presented a pipeline to detect BBH gravitational wave events in LIGO data developed by training neural networks on generated signals injected into real noise. With an ensemble of ten models trained with different characteristics, we have performed a full search on all of the two-detector LIGO data in O1, O2, and O3 for BBH events. 

While our pipeline fails to detect 18 cataloged BBH events, we do successfully detect 57, along with 57 false positives (noise events at least as significant as the least significant true positive).
By performing parameter inference, this greatly reduced number of false positive event candidates can be ruled out.
Out of the true positives that we failed to detect, they can all be explained as being caused by either low SNR in one detector, a low network SNR, or a low chirp mass.
Therefore, outside of those conditions, our pipeline detects all of the cataloged BBH events.

Furthermore, we have performed a significant test of our pipeline by evaluating it on over 100 years of time-shifted data to characterize the False Alarm Rate (FAR), and thus assign a FAR to each candidate.

This research represents an important early step in developing neural network-based searches of gravitational waves, and a crucial test of their application to multiple entire observing runs of data. A key next step for this pipeline will be testing it on LIGO's fourth observing run O4, when it is complete, which has already brought the number of detected GW events into the hundreds.

There are several key steps to further develop our pipeline: first, it can be extended to detect BNS and NSBH events, which are very important to detect for electromagnetic follow-up, though it will likely be difficult. The pipeline can be extended to work with a variable number of detectors to include Virgo and KAGRA, and still work for times without all detectors available, albeit at less sensitivity. Finally, this neural network pipeline can be extended to classify the type of GW event and give estimates of the event parameters like chirp mass and sky location. This approach has the potential to greatly speed up parameter estimation and follow-up observations compared to current methods.

\section*{Acknowledgments}

This research has made use of data or software obtained from the Gravitational Wave Open Science Center (gwosc.org), a service of the LIGO Scientific Collaboration, the Virgo Collaboration, and KAGRA. This material is based upon work supported by NSF's LIGO Laboratory which is a major facility fully funded by the National Science Foundation, as well as the Science and Technology Facilities Council (STFC) of the United Kingdom, the Max-Planck-Society (MPS), and the State of Niedersachsen/Germany for support of the construction of Advanced LIGO and construction and operation of the GEO600 detector. Additional support for Advanced LIGO was provided by the Australian Research Council. Virgo is funded, through the European Gravitational Observatory (EGO), by the French Centre National de Recherche Scientifique (CNRS), the Italian Istituto Nazionale di Fisica Nucleare (INFN) and the Dutch Nikhef, with contributions by institutions from Belgium, Germany, Greece, Hungary, Ireland, Japan, Monaco, Poland, Portugal, Spain. KAGRA is supported by Ministry of Education, Culture, Sports, Science and Technology (MEXT), Japan Society for the Promotion of Science (JSPS) in Japan; National Research Foundation (NRF) and Ministry of Science and ICT (MSIT) in Korea; Academia Sinica (AS) and National Science and Technology Council (NSTC) in Taiwan. 

The computations in this paper were run
on the FASRC Cannon cluster supported by the FAS Division of Science Research Computing Group at Harvard
University.

This work is supported by the National Science Foundation under Cooperative Agreement PHY-2019786 (The NSF AI Institute for Artificial Intelligence and Fundamental Interactions, \url{http://iaifi.org/}).

\bibliography{bibliography}
\clearpage
\appendix
\section{Pipeline Output Details}

Table \ref{tab:full_output} shows the output of our model pipeline on each of the cataloged BBH events in O1--O3 with two detectors of data. The table includes the most important parameters for detection, the Network SNR and Chirp Mass, the FAR assigned by LVC and our pipeline, and finally whether we successfully detected the event.

\onecolumngrid
\begin{longtblr}[
  caption = {Output of our model pipeline on each of the cataloged BBH events in O1--O3. The FARs are given per year. The limits on our Model FAR are due to the size of the 100 year time-shifted search.},
  label = {tab:full_output},
]{
  colspec = {ccccccc},
    rowhead = 1,
    rowfoot = 0,
    hline{1} = {solid},
    hline{Z} = {solid},
}
\hline
Event Name & Network SNR & Chirp Mass (M$_\odot$) & LVC FAR & Model Significance & Model FAR & Detected \\
\hline
GW150914 & 26.0 & 28.6 & $<10^{-7}$ & 12.83 & $<0.01$ & True \\
GW151012 & 10.0 & 15.2 & $<10^{-2}$ & 2.75 & 0.04 & True \\
GW151226 & 13.1 & 8.9 & $<10^{-7}$ & 1.23 & 12.14 & True \\
GW170104 & 13.8 & 21.4 & $<10^{-7}$ & 5.38 & $<0.01$ & True \\
GW170729 & 10.8 & 35.4 & 0.02 & 4.26 & $<0.01$ & True \\
GW170809 & 12.8 & 24.9 & $<10^{-7}$ & 3.43 & 0.01 & True \\
GW170814 & 17.7 & 24.1 & $<10^{-7}$ & 6.32 & $<0.01$ & True \\
GW170818 & 12.0 & 26.5 & $<10^{-5}$ & 0.31 & $>62.29$ & False \\
GW170823 & 12.2 & 29.2 & $<10^{-7}$ & 3.26 & 0.02 & True \\
GW190403\_051519 & 7.6 & 34.0 & 7.7 & 0.15 & $>62.29$ & False \\
GW190408\_181802 & 14.6 & 18.5 & $<10^{-5}$ & 6.21 & $<0.01$ & True \\
GW190412\_053044 & 19.8 & 13.3 & $<10^{-5}$ & 6.49 & $<0.01$ & True \\
GW190413\_052954 & 9.0 & 24.5 & 0.82 & 1.29 & 10.14 & True \\
GW190413\_134308 & 10.6 & 33.3 & 0.18 & 2.11 & 0.67 & True \\
GW190421\_213856 & 10.7 & 31.4 & $<10^{-8}$ & 3.94 & $<0.01$ & True \\
GW190426\_190642 & 8.7 & 76.0 & 4.1 & 0.44 & $>62.29$ & False \\
GW190503\_185404 & 12.2 & 29.3 & $<10^{-5}$ & 5.28 & $<0.01$ & True \\
GW190512\_180714 & 12.7 & 14.6 & $<10^{-5}$ & 2.46 & 0.13 & True \\
GW190513\_205428 & 12.5 & 21.8 & $<10^{-5}$ & 6.89 & $<0.01$ & True \\
GW190514\_065416 & 8.0 & 29.1 & 2.8 & 1.78 & 2.08 & True \\
GW190517\_055101 & 10.8 & 26.5 & $<10^{-5}$ & 2.62 & 0.06 & True \\
GW190519\_153544 & 15.9 & 44.3 & $<10^{-5}$ & 3.96 & $<0.01$ & True \\
GW190521\_030229 & 14.3 & 63.3 & $<10^{-3}$ & 2.35 & 0.21 & True \\
GW190521\_074359 & 25.9 & 32.8 & $<10^{-5}$ & 10.06 & $<0.01$ & True \\
GW190527\_092055 & 8.0 & 23.9 & 0.23 & 0.78 & 39.9 & True \\
GW190602\_175927 & 13.2 & 48.0 & $<10^{-5}$ & 3.44 & $<0.01$ & True \\
GW190701\_203306 & 11.2 & 40.2 & $<10^{-7}$ & 2.49 & 0.12 & True \\
GW190706\_222641 & 13.4 & 45.6 & $<10^{-5}$ & 6.04 & $<0.01$ & True \\
GW190707\_093326 & 13.1 & 8.4 & $<10^{-5}$ & 3.55 & $<0.01$ & True \\
GW190719\_215514 & 7.9 & 22.8 & 0.63 & 1.91 & 1.34 & True \\
GW190720\_000836 & 10.9 & 9.0 & $<10^{-5}$ & 0.23 & $>62.29$ & False \\
GW190725\_174728 & 9.1 & 7.4 & 0.46 & 0.61 & 55.65 & False \\
GW190727\_060333 & 11.7 & 29.4 & $<10^{-5}$ & 5.96 & $<0.01$ & True \\
GW190728\_064510 & 13.1 & 8.6 & $<10^{-5}$ & 3.01 & 0.04 & True \\
GW190731\_140936 & 8.8 & 29.7 & 0.33 & 1.38 & 7.72 & True \\
GW190803\_022701 & 9.3 & 27.6 & 0.07 & 1.8 & 1.94 & True \\
GW190805\_211137 & 8.1 & 31.9 & 0.63 & 0.05 & $>62.29$ & False \\
GW190828\_063405 & 16.5 & 24.6 & $<10^{-5}$ & 9.43 & $<0.01$ & True \\ 
GW190828\_065509 & 10.2 & 13.4 & $<10^{-5}$ & 1.59 & 3.73 & True \\
GW190915\_235702 & 13.1 & 24.4 & $<10^{-5}$ & 5.6 & $<0.01$ & True \\
GW190916\_200658 & 8.1 & 26.9 & 4.7 & 0.93 & 27.76 & True \\
GW190924\_021846 & 12.0 & 5.8 & $<10^{-5}$ & 0.54 & $>62.29$ & True \\
GW190926\_050336 & 8.1 & 24.4 & 1.1 & 1.93 & 1.29 & True \\
GW190929\_012149 & 9.7 & 35.6 & 0.16 & 2.26 & 0.34 & True \\
GW190930\_133541 & 9.7 & 8.5 & 0.01 & 0.98 & 24.35 & True \\
GW191103\_012549 & 8.9 & 8.34 & 0.46 & 0.56 & 60.58 & True \\
GW191105\_143521 & 9.7 & 7.82 & 0.01 & 0.96 & 25.96 & True \\
GW191109\_010717 & 17.3 & 47.5 & $<10^{-8}$ & 6.32 & $<0.01$ & True \\
GW191113\_071753 & 7.9 & 10.7 & 26.0 & 1.03 & 21.55 & False \\
GW191126\_115259 & 8.3 & 8.65 & 3.2 & 0.06 & $>62.29$ & False \\
GW191127\_050227 & 9.2 & 29.9 & 0.25 & 2.65 & 0.06 & True \\
GW191129\_134029 & 13.1 & 7.31 & $<10^{-5}$ & 3.04 & 0.02 & True \\
GW191204\_110529 & 8.9 & 19.8 & 3.3 & 0.24 & $>62.29$ & False \\
GW191204\_171526 & 17.4 & 8.56 & $<10^{-5}$ & 5.46 & $<0.01$ & True \\
GW191215\_223052 & 11.2 & 18.4 & $<10^{-5}$ & 3.9 & $<0.01$ & True \\
GW191222\_033537 & 12.5 & 33.8 & $<10^{-5}$ & 5.36 & $<0.01$ & True \\
GW191230\_180458 & 10.4 & 36.5 & 0.05 & 4.05 & $<0.01$ & True \\
GW200128\_022011 & 10.6 & 32.0 & $<10^{-3}$ & 2.9 & 0.04 & True \\
GW200129\_065458 & 26.8 & 27.2 & $<10^{-5}$ & 10.52 & $<0.01$ & True \\
GW200202\_154313 & 10.8 & 7.49 & $<10^{-5}$ & 0.08 & $>62.29$ & False \\
GW200208\_130117 & 10.8 & 27.7 & $<10^{-1}$ & 2.47 & 0.12 & True \\
GW200208\_222617 & 7.4 & 19.8 & 4.8 & 0.62 & 54.08 & False \\
GW200209\_085452 & 9.6 & 26.7 & 0.05 & 3.13 & 0.02 & True \\
GW200210\_092254 & 8.4 & 6.56 & 1.2 & 0.16 & $>62.29$ & False \\
GW200216\_220804 & 8.1 & 32.9 & 0.35 & 1.03 & 21.32 & False \\
GW200219\_094415 & 10.7 & 27.6 & $<10^{-9}$ & 1.96 & 1.16 & True \\
GW200220\_061928 & 7.2 & 62.0 & 6.8 & 0.01 & $>62.29$ & False \\
GW200220\_124850 & 8.5 & 28.2 & 30.0 & 1.05 & 20.47 & True \\
GW200224\_222234 & 20.0 & 31.1 & $<10^{-5}$ & 9.42 & $<0.01$ & True \\
GW200225\_060421 & 12.5 & 14.2 & $<10^{-5}$ & 4.64 & $<0.01$ & True \\
GW200306\_093714 & 7.8 & 17.5 & 24.0 & 0.94 & 26.86 & False \\
GW200308\_173609 & 4.7 & 34.0 & 2.4 & 0.04 & $>62.29$ & False \\
GW200311\_115853 & 17.8 & 26.6 & $<10^{-5}$ & 7.55 & $<0.01$ & True \\
GW200316\_215756 & 10.3 & 8.75 & $<10^{-5}$ & 0.19 & $>62.29$ & False \\
GW200322\_091133 & 4.5 & 15.0 & 140.0 & 0.38 & $>62.29$ & False \\
\hline
\end{longtblr}

\end{document}